
\magnification=\magstep1
\raggedbottom

\font\bo=cmbx10  scaled\magstep1
\font\tsc=cmcsc10

\hsize=15.truecm  \hoffset=1.truecm
\vsize=23.truecm  \voffset=-.5truecm
\topskip=1.truecm \leftskip=0.truecm

\pageno=1
\footline={\hfill}
\headline={\ifnum\pageno=1\hfil\else\hss\tenrm-\ \folio\ -\hss\fi}

\let\cl\centerline  \let\rl\rightline   \let\ll\leftline
\let\bs\bigskip    \let\ss\smallskip

\null
\rl{hep-ph/9403245}
\rl{March 1994}
\vskip2.truecm
{\bo
\cl{SCALING LAWS IN}\ss
\cl{HIERARCHICAL CLUSTERING MODELS}\ss
\cl{WITH POISSON SUPERPOSITION}
}

\vskip1.5truecm
\cl{\tsc S. Hegyi}

\footnote\ {\it hegyi@rmki.kfki.hu}

\baselineskip=11pt
{\it
\cl{KFKI Research Institute}
\cl{for Particle and Nuclear Physics}
\cl{of the Hungarian Academy of Sciences,}
\cl{H--1525 Budapest 114, P.O. Box 49. Hungary}
}

\vfill

\baselineskip=12pt
\midinsert\narrower\narrower
\noindent
{\rm
\underbar{ABSTRACT}\hskip.3truecm
Properties of cumulant- and combinant ratios are studied for
multihadron final states composed of Poisson distributed clusters.
The application of these quantities to ``detect'' clusters is
discussed. For the scaling laws which hold in hierarchical
clustering models (void scaling, combinant scaling) a generalization
is provided. It is shown that testing hierarchical models is
meaningful only for phase-space volumes not larger than the
characteristic correlation length introduced by Poisson superposition.
Violation of the scaling laws due to QCD effects is predicted.

}
\endinsert

\vfill\eject

\baselineskip=13pt
\parskip=10pt

\noindent
{\bo 1. Introduction}

\noindent
In galaxy clustering studies there is accumulating
observational evidence in favour of the so-called hierarchical
models (see e.g. refs.~[1-5] and ref.~[6] for a historical account).
Initiated by the work of Carruthers and Sarcevic [7] and the Tucson
group [8,9] considerable interest has been devoted to
this subject over the last few years also in studying the nature of
correlations of multihadron
final states [10-13]. In the field of multiparticle dynamics one speaks
about Linked Pair Approximation (LPA) because of the special hierarchy
of the higher-order correlation functions.
Throughout this paper we shall use both naming conventions.

In the framework of the LPA
the two-particle cumulant correlations, defined in terms of the
inclusive single- and two-particle density correlations
$\rho_1$ and $\rho_2$ as
$$
    {\cal K}_2(1,2)=\rho_2(1,2)-\rho_1(1)\rho_1(2),\eqno(1.1)
$$
provide the building blocks of the higher-order
cumulants~[7,8]. In eq.~(1.1) the arguments denote coordinates e.g.
on the rapidity axis.
The cumulants(densities) are known also as irreducible(reducible)
correlations since the lower-order background correlation terms involved
by $\rho_q$ are subtracted in ${\cal K}_q$.
The cumulants measure genuine
$q$-particle correlations, the degree of independence of their arguments.
In the case of full statistical independence
(each argument is independent of all the others) the $q$-particle
density $\rho_q$ factorizes into the product of $q$ single-particle
densities and the $q$-particle cumulant
${\cal K}_q$ vanishes. In the LPA the normalized cumulants
$$
    \kappa_q(1,...,q)={\cal K}_q(1,...,q)/
    \rho_1(1)...\rho_1(q)\eqno(1.2)
$$
are built up as sums of products of linked two-particle normalized
cumulants. For example, $\kappa_3$ is composed according to
$$
\eqalign{
    \kappa_3(1,2,3)={A_3\over3}\,\big[
    \kappa_2(1,2)\kappa_2(2,3)&+
    \kappa_2(2,3)\kappa_2(3,1)\cr&+
    \kappa_2(3,1)\kappa_2(1,2)\big]
}\eqno(1.3)
$$
where the constant $A_3$ is a free parameter to be determined (by
definition, $A_1=A_2=1$). In the astrophysical
literature the constants $A_q$
are known as hierarchical amplitudes. The Linked Pair Approximation
is in fact a special case of the hierarchical
models~[11]. Some types of topologically distinct graphs
allowed in the construction of higher-order galaxy correlation
functions are absent in the LPA.
These are the so-called non-snake graphs. In lowest order the two
schemes are equivalent, leading to eq.~(1.3), and they coincide at the
level of the integrated cumulants.

Two years before the discovery of the
successfulness of eq.~(1.3) for galaxy correlation data~[2]
Mandelbrot arrived at the same functional form by
constructing a simple model for galaxy clustering~[14].
In the model the positions of galaxies are stepping points of a
Rayleigh-L\'evy random walk which gives rise to an unbounded fractal
galaxy distribution. Later it was realized that the distribution of
galaxies cannot be a pure, unbounded fractal; it contradicts to the
fine structure of the observed galaxy distribution~[6]. The model was
modified by introducing many Rayleigh-L\'evy random walk fractals
distributed according to a Poisson process~[1]. The Poisson
superposition of bounded fractal clumps well reproduces the visual
appearance of the galaxy distribution in the sky and it is in
agreement with the hierarchical structure of higher-order correlation
functions [1,6].

According to recent work of Giovannini, Lupia and Ugoccioni
the hierarchical structure of cumulant correlations and
the Poisson superposition principle play an important
role in multiparticle dynamics too. The Torino group studies the
nature of inside-jet correlations in $e^+e^-$ annihilations
using {\tsc Jetset} 7.2 Monte Carlo events in a wide c.m. energy
range up to \hbox{$\sqrt s=1000$} GeV [13, 15-17].
One of the main results is the identification of jets originated
by a quark(antiquark) and a gluon. The correlations are in agreement
with the hierarchical ansatz in quark- and gluon-jets.
Moreover, it is found that both types of jets
consist of Poisson distributed clumps of particles called clans.
Within gluon-jets the correlations are stronger (the clans have larger
particle content) which is attributed to QCD effects. In gluon initiated
jets the dominant mechanism is the self-interaction of gluons whereas
quark-jets are controlled by gluon bremsstrahlung emission~[13].

The Poisson superposition introduces a characteristic
correlation length in hierarchical models.
The main goal of the present paper is the study
of the consequences of this newly appearing length scale.
Although correlations of multihadron final states will be
investigated some of the results may have applications in other fields,
particularly in galaxy clustering studies.

\bs
\noindent
{\bo 2. Basic definitions}

\noindent
Currently available data for hadrons and galaxies do not allow precise
determination of the higher-order correlation functions. But testing
the validity of hierarchical models is possible through the
integrated cumulants that provide the factorial cumulant moments of
the underlying count distributions. Moments can be determined up to
$8th$ order for galaxy catalogs~[4,5]
and up to $5th$ order for multihadron final states~[8].
In this section we collect some basic definitions and formulae
concerning count distributions and various moments. For more details,
see e.g. refs.~[10] and [18].

Let us start with the generating function for the multiplicity- or count
distributions $P_n$, conveniently defined by
$$
    {\cal G}(z)=\sum_{n=0}^\infty P_nz^n.\eqno(2.1)
$$
The connection with the density correlation functions $\rho_q$ is
given by the power expansion of ${\cal G}(z)$:
$$
    {\cal G}(z)=1+\sum_{q=1}^\infty {(z-1)^q\over q!}\,\xi_q\eqno(2.2)
$$
where the $\xi_q$ are the integrated densities providing the factorial
moments of $P_n$. The integrated cumulants ${\cal K}_q$, being the
factorial cumulant moments $f_q$ of $P_n$, are defined by the power
expansion of the logarithm of ${\cal G}(z)$:
$$
    \ln{\cal G}(z)=\sum_{q=1}^\infty {(z-1)^q\over q!}f_q.\eqno(2.3)
$$
In the analysis of various count distributions occuring in nature
the so-called infinitely divisible distributions play a distinguished
role. In multiparticle dynamics their importance was emphasized by
Giovannini and Van Hove~[19,20]. A {\it discrete\/} distribution is
said to be infinitely divisible if its generating function has the
property that for all $k>0$ integer $\root k\of{{\cal G}(z)}$ is again
the generating function of a certain distribution~[21]. ${\cal G}(z)$
satisfies this property if and only if
${\cal G}(1)=1$ and
$$
    \ln{\cal G}(z)=\ln{\cal G}(0)+\sum_{q=1}^\infty C_qz^q.\eqno(2.4)
$$
In eq.~(2.4) the $C_q$ are the combinants of Gyulassy and
Kauffmann [22,23]. The combinants should obey
$C_q\geq0$ and
$$
    \sum_{q=1}^\infty C_q=-\ln{\cal G}(0)<\infty\eqno(2.5)
$$
for infinitely divisible distributions.
{}From eq.~(2.5) one sees that
the probability of detecting no particles at all in a certain
phase-space volume, the so-called void probability, is
$P_0={\cal G}(0)>0$ for discrete distributions satisfying the
conditions of infinite divisibility.
The generating function takes the form
$$
    {\cal G}(z)=\exp\bigg(\sum_{q=1}^\infty C_q(z^q-1)\bigg)\eqno(2.6)
$$
in terms of combinants.
The $C_q$ can be expressed as combinations of the
count probability ratios ${\cal P}_q=P_q/P_0$,
$$
    C_q={\cal P}_q-{1\over q}\sum_{r=1}^{q-1} r\,C_r{\cal P}_{q-r},
    \eqno(2.7)
$$
hence their name.
Eq.~(2.7) shows that the knowledge of $C_q$
requires only a finite number of count
probabilities. Furthermore, we need not know the probabilities
themselves: the combinants follow directly from the unnormalized
topological cross sections since they involve only ratios of
probabilities.
It is also seen from eq.~(2.7) that in general the $C_q$ can take
negative values as well for $q\geq2$. In this case a necessary
condition of the infinite divisibility of $P_n$ is not satisfied.
The combinants have some advantageous features, e.g. they share
common properties with the cumulant moments. For further details
see refs.~[23,24].

\bs\bs
{\bo
\ll{3. Cumulant- and combinant ratios in}
\ll{\phantom{3. }Poisson cluster models}
}

\noindent
In order to study hierarchical models with Poisson superposition we
shall utilize the so-called Poisson cluster models. These are closely
related to the infinitely divisible distributions discussed in the
previous section. An advantageous feature of these models is the
absence
of free parameters. The clan production picture of hadronization
developed by Giovannini and Van Hove~[19,20] is a well known
example of Poisson cluster models. We start with the basic equations.

Assume that the observed events (the phase-space distribution of
multihadron final states produced in a certain collision process
or the galaxy distribution in the sky as a single ``event'') can be
decomposed into Poisson distributed clusters.
In the Poisson cluster models the generating function of the
total event multiplicity distribution $P_n$ takes the form
$$
\eqalign{
    {\cal G}(z)&=
    \exp\left(\bar{\cal C}({\cal H}(z)-1)\right)\cr
    &={\cal P}\big({\cal H}(z)\big)
}\eqno(3.1)
$$
where ${\cal P}(z)$ and ${\cal H}(z)$ stand for the generating
functions of the Poissonian cluster distribution
and the arbitrary distribution of particles inside the
clusters. ${\cal G}(z)$ is obtained as the
convolution of ${\cal P}(z)$ and ${\cal H}(z)$. Discrete
distributions
having a generating function of the above form are known as compound
Poisson distributions~[21]. Eq.~(3.1) is another
way of writing ${\cal G}(z)$ for infinitely divisible
distributions, eq.~(2.6), with
\hbox{$\bar{\cal C}=-\ln{\cal G}(0)$} and
$$
    {\cal H}(z)=1-{\ln{\cal G}(z)\over\ln {\cal G}(0)}=
    \sum_{q=1}^\infty p_qz^q.\eqno(3.2)
$$
In eq.~(3.2) the the $q$-particle count probability within a single
cluster, $p_q$, is found to be~[24,25]
$$
     p_q={C_q\over\sum_q C_q}\,,\eqno(3.3)
$$
reminiscent of the definition of count probabilities $P_n$
in terms of the topological cross sections $\sigma_n$,
\hbox{$P_n=\sigma_n/\sum_n\sigma_n$}.

Since ${\cal H}(0)=0$ each cluster must contain at least one particle,
i.e. $p_0=C_0~=~0$.
Accordingly, ${\cal G}(0)={\cal P}(0)$ and the
probability of detecting no particles, $P_0$, provides also
the probability of detecting no clusters in a certain phase-space
volume. Hence the two basic parameters of the model, the average
cluster multiplicity, $\bar{\cal C}$, and the average multiplicity
within the clusters, $\bar q$, are found to be~[12,26]
$$
    \bar{\cal C}=-\ln P_0\qquad\hbox{\rm and}\qquad
    \bar q=\bar n/\bar{\cal C}\eqno(3.4)
$$
with $\bar n=f_1=\xi_1$ being the total event average multiplicity.
It is worth noticing that eq.~(2.6) can be written also in the form
$$
    {\cal G}(z)=\prod_{q=1}^\infty\exp\big(C_q(z^q-1)\big),\eqno(3.5)
$$
i.e. as the product of the generating function of
a Poisson distribution of particle singlets having
mean $C_1$, a Poisson distribution of particle pairs having mean
$C_2$, and so on~[21]. Thus a formally equivalent
derivation of the Poisson cluster
models is possible in which the total events are composed of Poisson
distributed particle \hbox{$q$-tuples} with average multiplicities
$C_q$ instead of identical clusters or
clans. This view is closely reminiscent of the Mayer cluster
expansion of statistical mechanics. Recall that the $C_q$ can take
negative values as well for $q\geq2$. In this case the interpretation
of combinants as unnormalized count probabilities within identical
Poisson clusters or as the average multiplicities of Poisson
distributed particle $q$-tuples loses its meaning.

On the basis of eq.~(3.1) let us consider
the relationship between quantities corresponding
to the observed events and quantities at the level of the individual
clusters. For the total event factorial cumulants we get~[27]
$$
\eqalign{
    f_q&={d^q\over dz^q}\ln{\cal G}(z)\,\Big|_{\,z=1}\cr
       &=\,\bar{\cal C}\,{d^q\over dz^q}{\cal H}(z)\,\Big|_{\,z=1}
        =\bar{\cal C}\zeta_q\cr
}\eqno(3.6)
$$
with $\zeta_q$ denoting the factorial moments within a single cluster.
As obtained in eq.~(3.3),
the combinants of the total events are related to the
count probabilities inside the clusters:
$$
\eqalign{
    C_q&={1\over q!}{d^q\over dz^q}\ln{\cal G}(z)\,\Big|_{\,z=0}\cr
       &=\,\bar{\cal C}\,{1\over q!}{d^q\over dz^q}
       {\cal H}(z)\,\Big|_{\,z=0}=\bar{\cal C}p_q.\cr
}\eqno(3.7)
$$
We see that ratios of quantities derived from $\ln{\cal G}(z)$ are
equivalent to ratios of quantities derived from ${\cal H}(z)$ and
characterize the individual clusters. This feature is well
suited to ``detect''
clusters. When both $p$- and $q$-particle correlations exist
any of the ratios
$f_q/f_p$, $C_q/C_p$ or $C_q/f_p$ should become
{\it independent\/} of
changes in the size of the phase-space volume
if a certain characteristic correlation length is exceeded.
Cuts in phase-space larger than the typical cluster size
correspond to cuts in the number of clusters and do not influence
single cluster statistics.
Therefore a flattening is expected in the large-scale behaviour
of the above quantities, provided that translation invariance holds
for extended volumes.
Flattening of the ratios $f_q/f_1$ for wide rapidity intervals
has already been studied by Dias de Deus using the assumption that
multihadron final states consist of a fixed number of
uncorrelated clusters~[28].
In this case the constancy follows from the additivity property
of the cumulant moments. The large binsize ($\delta y\geq2$)
plateau in the behaviour of the $f_q/f_1$ was indeed observed
for NA22
data at $\sqrt s=22$ GeV and for UA5 data at $\sqrt s=546$
GeV~[28,29].

\bs\bs
{\bo
\ll{4. Hierarchical models, Poisson}
\ll{\phantom{4. }superposition and scaling}
}

\noindent
In this section we consider properties of the
cumulant- and combinant ratios
again but the Poisson superposition principle
is combined with the simultaneous validity of the
Linked Pair Approximation. Assuming translation invariance and
the linked-pair hierarchy of the higher-order cumulant
correlation functions the
strip-integration~[8] of $\kappa_q$ produces normalized
factorial cumulant moments obeying the recurrence relation
$$
    K_q\equiv{f_q\over f_1^q}=A_qK_2^{q-1}.\eqno(4.1)
$$
The hierarchical amplitudes $A_q$ should be
independent of collision energy and phase-space volume.
Eq.~(4.1) is valid also for hierarchical models that allow not only
snake-type graphs in the construction of higher-order cumulant
correlation functions. The integrated amplitudes in eq.~(4.1)
may deviate from
the local ones appearing in the expressions for $\kappa_q$.
In addition to the direct
test of eq.~(4.1) it is possible to check the validity of the LPA
through various {\it scaling laws\/}.
For the ratio of two factorial cumulants eq.~(4.1) yields
$$
    {f_q\over f_p}={A_q\over A_p}(f_2/f_1)^{q-p}.\eqno(4.2)
$$
Expressing the combinants in terms of factorial cumulant moments
[24,25,30],
$$
    C_q=\sum_{r=q}^\infty{r\choose q}\,{(-1)^{r-q}\over r!}\,f_r,
    \eqno(4.3)
$$
one sees that the cumulant- and combinant ratios
considered in the previous section obey
a universal behaviour in hierarchical models.
Any of these quantities should depend on collision energy and
phase-space volume {\it not arbitrarily\/} but only through
the combination $f_2/f_1$. Beside eq.~(4.2) we have e.g.
$$
\eqalign{
    {C_q\over f_p}&=
    \sum_{r=q}^\infty{r\choose q}{(-1)^{r-q}\over r!}
    {A_r\over A_p}(f_2/f_1)^{r-p}\cr&\equiv
    \chi_{qp}(f_2/f_1),\cr
}\eqno(4.4)
$$
the combinant-to-cumulant ratios chosen at fixed $q$, $p$ from
different energies and volumes scale to a universal curve when
plotted against $f_2/f_1$. In eq.~(4.4)
the contribution of an infinite number of factorial cumulant moments
is confined into the scaling variable $f_2/f_1$.
The $\chi_{qp}$ provide the generalization of the scaling functions
introduced in ref.~[25] which are the $p=1$ special cases of
eq.~(4.4).

The generalization of the famous void scaling function [31,32]
can be obtained in a similar manner. The probability of detecting no
particles in a certain  phase-space volume obeys the scaling law
$$
\eqalign{
    {-\ln P_0\over f_p}&=\sum_{r=1}^\infty
    {(-1)^{r-1}\over r!}{A_r\over A_p}(f_2/f_1)^{r-p}\cr&\equiv
    \chi_{0p}(f_2/f_1)\cr
}\eqno(4.5)
$$
as is seen from eq.~(2.3) by substituting $z=0$.
The zero subscript of $\chi$ in eq.~(4.5)
refers to the property that these scaling
functions involve only the void probability.
Through eq.~(2.5) we get $\chi_{0p}=\sum_{q=1}^\infty\chi_{qp}$.
Similarly to the void scaling functions, the $\chi_{qp}$
constructed from the combinants characterize the distribution of
empty regions in phase-space. The $C_q$ can be interpreted as
minus the logarithm of the
probability that a certain phase-space volume is empty of particle
$q$-tuples, see eq.~(3.5).  The advantage of $\chi_{qp}$ over
$\chi_{0p}$ lies in the fact that the contribution of low-order
factorial cumulants can be excluded in a systematic manner
by increasing $q$ (observe
that $f_1$ and $f_2$ involved by $P_0$ carry no information on the
validity of the LPA).

Since the relationship between count probabilities and factorial
moments is the same as the relationship between
combinants and factorial cumulants [23,30] a scaling law analogous
to eq.~(4.4) holds for the count probabilities
$P_n$ if the recurrence relation
$$
    F_q\equiv{\xi_q\over\xi_1^q}=A_qF_2^{q-1}\eqno(4.6)
$$
with constant coefficients $A_q$
is satisfied for the normalized factorial moments:
$$
\eqalign{
    {P_n\over\xi_p}&=
    \sum_{r=n}^\infty{r\choose n}{(-1)^{r-n}\over r!}
    {A_r\over A_p}(\xi_2/\xi_1)^{r-p}\cr
    &\equiv\eta_{np}(\xi_2/\xi_1)\cr
}\eqno(4.7)
$$
with $n$, $p\geq1$ since $\xi_0=1$.
That is to say, if one plots the ratios $P_n/\xi_p$ chosen at fixed
$n$, $p$ against the combination $\xi_2/\xi_1$ the validity of
eq.~(4.6) results in a universal curve,
\hbox{$\eta_{np}(\xi_2/\xi_1)$}, instead of many different behaviours
corresponding to different collision energies and volume sizes.
Count probability ratios not
involving $P_0$ constitute another set of scaling functions.
The analogue of void scaling for eq.~(4.6) is provided by
\hbox{$\eta_{0p}\equiv(1-P_0)/\xi_p=\sum_{n=1}^\infty\eta_{np}$}.

Eq.~(4.6) characterizes
monofractal density fluctuations. They are expected to occur e.g.
in second-order phase transitions. Hence the scaling law of the
multiplicity distributions expressed by eq.~(4.7) can in principle
signal the formation of quark-gluon plasma
if QCD undergoes a second-order transition~[30].
It should be emphasized that eq.~(4.6) holds for any field theory
with {\it dimensionless\/} coupling~[33]. For example, in QCD
with fixed coupling constant $\alpha_{\rm s}$ the density
fluctuations
are concentrated into randomly distributed monofractal patches of
phase-space~[34]. With running coupling $\alpha_{\rm s}$ multifractal
fluctuations appear and the exponent on the rhs. of
eq.~(4.6) gets multiplied by a $q$-dependent factor.
QCD effects violate the above scaling rule of the count probabilities
$P_n$.

We turn our attention to
the validity of the LPA and the resulting scaling laws
if the observed events are composed of Poisson distributed clusters.
First of all, from eq.~(3.6) we get
${\cal F}_q=\bar{\cal C}^{q-1}K_q$ and
$$
    A_q\equiv{K_q\over K_2^{q-1}}={{\cal F}_q\over{\cal F}_2^{q-1}}
    \eqno(4.8)
$$
with ${\cal F}_q$ denoting the normalized factorial
moments within a single cluster, $\zeta_q/\zeta_1^q$.
One sees that the hierarchical amplitudes
$A_q$ are equivalent to single cluster statistics.
Consequently, they should be independent of changes in the size of
the phase-space volume
if the typical cluster size is exceeded,
{\it regardless\/} of the validity of the LPA.
For the scaling laws we have similar equivalence relations.
According to eqs.~(3.6) and (3.7) the scaling functions of eq.~(4.4)
at the level of the observed events
are equivalent to the scaling functions of eq.~(4.7)
at the level of the individual clusters:
$$
    \chi_{qp}(f_2/f_1)\,\big|_{\,\rm total\ event}
    =\,
    \eta_{qp}(\xi_2/\xi_1)\,\big|_{\,\rm single\ cluster}.\eqno(4.9)
$$
Moreover, the void scaling functions of eq.~(4.5) are equivalent
to $1/\zeta_q$, the reciprocal of the unnormalized
factorial moments inside the clusters.
Thus for multihadron final states composed of Poisson distributed
clusters testing eq.~(4.1) and the scaling laws expressed by
eqs.~(4.4) and (4.5) is meaningful only for phase-space
volumes not exceeding the characteristic correlation length
introduced
by Poisson superposition. Cuts in phase-space larger than the
typical cluster size are equivalent to cuts
in the number of clusters and leave single cluster statistics
unchanged. We mention that the cluster size
may vary with energy, e.g. in $hh$ collisions the threshold binsize
in rapidity at which the constancy of the ratios $f_q/f_1$ appears
is increasing from $\delta y\approx2$ to $\delta y\approx3$ between
$\sqrt s=22$ and $546$~GeV [28,29]. In the $\chi$-scaling analyses
already performed [11-13] the chosen cuts in rapidity do not exceed
the above values.

Closing this section let us reconsider the promising results of the
Torino group mentioned in the Introduction. The finding that quark-
and gluon-jets consist of Poisson distributed groups of particles,
the clans, and the confirmation of hierarchical inside-jet
correlations
provide a direct connection to our results. The hierarchical
pattern
of cumulants in quark- and gluon-jets was deduced from the validity
of the void scaling law eq.~(4.5) for $p=1$~[13].
Due to the Poisson superposition this
scaling rule should follow from the validity of eq.~(4.6) at the
level of the individual clans. Since clan properties inside
gluon originated jets are
controlled by the self-interaction of gluons, deviations are expected
from the void- and combinant scaling functions in gluon-jets.
For running QCD
coupling constant $\alpha_{\rm s}$ eq.~(4.6) is no longer valid
[33,34] and this violates the inside-clan scaling rules of count
probabilities and unnormalized factorial moments
being equivalent to the inside-jet combinant- and void
scaling laws. In the \hbox{$\sqrt s=1000$}
GeV c.m. energy range investigated by the Torino group the
self-interaction of gluons may cause observable scaling violation
for gluon initiated jets.
Deviations from the void scaling function $\chi_{01}$ are indeed
observed~[13] and they are attributed to the absence of translation
invariance. The separation of the two sources of scaling violation
would be of particular importance. According to preliminary results
of De Wolf~[35] the breakdown of scaling can be seen also in the
behaviour of the higher-order $\chi_{q1}$ combinant-to-cumulant
ratios for {\tsc Jetset} Monte Carlo events.

\bs
\noindent
{\bo 5. Summary}

\noindent
One of the common problems in the study of the very large and very
small length-scale phenomena is the formation of structures --- the
texture of multihadron final states in phase-space and the
distribution of galaxies in the sky. The assumption of randomly
superimposed objects (clusters, clans) with cumulant correlation
functions obeying hierarchical structure at the level of the total
matter distribution plays a distinguished role in multiparticle
physics and particularly in galaxy clustering studies. We have
investigated
the properties of cumulant- and combinant ratios for hierarchical
models
with Poisson superposition. According to our results the behaviour
of the ratios $f_q/f_p$, $C_q/C_p$ and $C_q/f_p$ among others
considerably simplifies with the two assumptions:

\item{\it i)} From eqs.~(3.6) and (3.7) one obtains that for Poisson
distributed clusters (clans) the above quantities are equivalent to
single cluster statistics. Consequently, they should be independent
of changes in the size of the phase-space volume if the characteristic
correlation length introduced by Poisson superposition is exceeded.
Cuts in phase-space larger than the typical cluster size correspond
to cuts in the number of clusters and do not influence single cluster
statistics. Therefore large-scale constancy of the cumulant- and
combinant ratios is expected at a fixed energy.

\item{\it ii)} The consequence of the LPA-relation, eq.~(4.1),
is that the cumulant- and
combinant ratios should depend on collision energy and phase-space
volume not arbitrarily but
only through the momentum combination $f_2/f_1$. One arrives
e.g. at the nontrivial scaling laws expressed by eqs.~(4.4) and (4.5).
These are generalizations of the scaling laws introduced in refs.~[31]
and [25]. According to {\it i)\/}
for a set of Poisson distributed clusters all of the
hierarchical statistics ($A_q$, $\chi_{qp}$, $\chi_{0p}$) are
equivalent to statistics characterizing the individual clusters.
Therefore testing the validity of eq.~(4.1) and the resulting
scaling laws is meaningful only for
volume sizes not larger than the typical size of the clusters.

\noindent
By eq.~(4.8)
we are led to the conclusion that in hierarchical models
with Poisson superposition the linked-pair hierarchy of higher-order
correlation functions manifests itself primarily at the level of the
individual clusters. Within a single cluster the density correlation
functions and factorial moments (the reducible statistics) obey the
hierarchical structure. The cumulant correlation functions and
factorial cumulant moments (the irreducible statistics) exhibit the
hierarchical pattern at the level of the total events as the
consequence of Poisson superposition. Eq.~(4.1) and the resulting
combinant- and void scaling laws hold
for any field theory with dimensionless coupling
if randomly superimposed clusters are present, e.g.
in QCD with fixed coupling constant $\alpha_{\rm s}$.
Restoration of the running of $\alpha_{\rm s}$ causes violation
of the scaling laws. This can be most significant
in $e^+e^-$ annihilations for gluon initiated jets controlled by
gluon self-interaction.

We close with a list of some measurements that may shed more light on
the validity of hierarchical models and Poisson superposition:

\item{$\bullet$} The Poisson cluster models discussed in the present
paper can be tested e.g. by measuring the $C_q$. Observation of
negative
combinants signals that the underlying count distributions are not
infinitely divisible. In this case the Poisson cluster
models are inappropriate to represent the data.

\item{$\bullet$} The presence of clusters can be revealed by
correlation length measurements. They consist of
determining the threshold volume size at which the constancy of
the cumulant- and combinant ratios appears. The threshold provides
an estimation for the typical size (e.g. rapidity extent) of the
Poisson distributed clusters.

\item{$\bullet$} Measurement of the higher-order combinant scaling
functions, e.g. the $\chi_{q1}$ for $q\geq3$, is important
to confirm eq.~(4.1).
These have the advantage over the void scaling function $\chi_{01}$
that the contribution of $f_1$
and $f_2$ is excluded. The first two factorial cumulants have
considerable influence on the shape of $\chi_{01}$ but they
carry no information on the validity of eq.~(4.1).

\item{$\bullet$} It would be interesting to see how clear is the
violation of $\eta$-scaling if
\hbox{$\chi$-scaling} holds valid. This provides
information on the selective power of scaling laws that have
different origin. The simplest possibility is the comparison of
$-\ln P_0/f_1$ and $(1-P_0)/\xi_1$ as functions of $f_2/f_1$ and
$\xi_2/\xi_1$ respectively.

\noindent
Lack of translation invariance
may considerably influence the proposed
measurements. Estimation of its significance is important to gain
reliable information on higher-order correlations obeying
hierarchical structure with the possible presence of clusters
distributed according to a Poisson process.
On the basis of the experimental data that will be available at
LEP200  and TEVATRON, at twice as large c.m. energies as present,
we have the chance of answering the open questions in the near
future.

\bs\bs\bs

\noindent
{\bo Acknowledgements}

\noindent
I am indebted to P. Carruthers for the many ways of inspiration.
This research was supported by the Hungarian Science Foundation under
grants No. OTKA-2972 and OTKA-F4019.

\vfill\eject
\parskip=0pt
\noindent
{\bo References}
\vskip.5cm
\frenchspacing

\item{[1]}  P.J.E. Peebles, {\it The Large-Scale Structure of the
            Universe\/}, \hfill\break
            Princeton University Press, 1980.\ss
\item{[2]}  E.J. Groth and P.J.E. Peebles, {\it Ap. J.\/}
            {\bf 217} (1977) 285.\ss
\item{[3]}  J.N. Fry and P.J.E. Peebles, {\it Ap. J.\/}
            {\bf 221} (1978) 19.\ss
\item{[4]}  I. Szapudi, A.S. Szalay and P. Bosch\'an,
            {\it Ap. J.\/} {\bf 390} (1992) 350.\ss
\item{[5]}  A. Meiksin, I. Szapudi and A.S. Szalay,
            {\it Ap. J.\/} {\bf 394} (1992) 87.\ss
\item{[6]}  P.J.E. Peebles, {\it Physica D\/}
	    {\bf 38} (1989) 273.\ss
\item{[7]}  P. Carruthers and I. Sarcevic,
            {\it Phys. Rev. Lett.\/} {\bf 63} (1989) 1562.\ss
\item{[8]}  P. Carruthers, H.C. Eggers, Q. Gao and
            I. Sarcevic,\hfill\break
            {\it Int. J. Mod. Phys.\/} {\bf A6} (1991) 3031.\ss
\item{[9]}  H.C. Eggers, P. Carruthers, P. Lipa and I. Sarcevic,
            \hfill\break {\it Phys. Rev.\/} {\bf C44} (1991) 1975.\ss
\item{[10]} P. Carruthers, {\it Acta Phys. Pol.\/}
            {\bf B22} (1991) 931.\ss
\item{[11]} E.A. De Wolf, in {\it Fluctuations and Fractal
            Structure,\/}\hfill\break
            Proc. of the Ringberg Workshop on Multiparticle
            Production, p. 222,\hfill\break ed. by R.C. Hwa, W. Ochs
	    and N. Schmitz, World Scientific, 1992.\ss
\item{[12]} S. Hegyi,
            {\it Phys. Lett.\/} {\bf B274} (1992) 214.\ss
\item{[13]} A. Giovannini, S. Lupia and R. Ugoccioni,
            {\it Z. Phys. C\/} {\bf 59} (1993) 427.\ss
\item{[14]} B. Mandelbrot, {\it Comptes Rendus (Paris)\/}
            {\bf 280A} (1975) 1551.\ss
\item{[15]} R. Ugoccioni, A. Giovannini and S. Lupia,
            {\it ``Void Analysis and Hierarchical \hfill\break
            Structure for Single Jets''\/},
            Torino Univ. Report, DFTT 41/93.\ss
\item{[16]} S. Lupia, A. Giovannini and R. Ugoccioni,
            {\it ``Hierarchical Structures \hfill\break
            in High Energy Collisions''\/},
            Torino Univ. Report, DFTT 51/93.\ss
\item{[17]} A. Giovannini, S. Lupia and R. Ugoccioni,
            {\it ``Void Analysis in \hfill\break
            Multiparticle Production''\/},
            Torino Univ. Report, DFTT 59/93.\ss
\item{[18]} P. Carruthers and C.C. Shih,
            {\it Int. J. Mod. Phys.\/} {\bf A2} (1987) 1447.\ss
\item{[19]} A. Giovannini and L. Van Hove,
            {\it Acta Phys. Pol.\/} {\bf 19} (1988) 495.\ss
\item{[20]} L. Van Hove and A. Giovannini,
            {\it Acta Phys. Pol.\/} {\bf 19} (1988) 917 and 931.\ss
\item{[21]} W. Feller, {\it Introduction to Probability Theory and
            its Applications\/}, \hfill\break Vol. I., Third Edition,
            Wiley 1971.\ss
\item{[22]} M. Gyulassy and S.K. Kauffmann,
            {\it Phys. Rev. Lett.\/} {\bf 40} (1978) 298.\ss
\item{[23]} S.K. Kauffmann and M. Gyulassy,
            {\it J. Phys. A\/} {\bf 11} (1978) 1715.\ss
\item{[24]} I. Szapudi and A.S. Szalay,
            {\it Ap. J.\/} {\bf 408} (1993) 43.\ss

\vfill\eject

\item{[25]} S. Hegyi,
            {\it Phys. Lett.\/} {\bf B309} (1993) 443.\ss
\item{[26]} S. Hegyi and T. Cs\"org\H o,
            {\it Phys. Lett.\/} {\bf B296} (1992) 256.\ss
\item{[27]} M. Le Bellac,
            {\it Acta Phys. Pol.\/} {\bf B4} (1973) 901.\ss
\item{[28]} J. Dias de Deus,
            {\it Phys. Lett.\/} {\bf B178} (1986) 301.\ss
\item{[29]} J. Dias de Deus,
            {\it Phys. Lett.\/} {\bf B185} (1987) 189.\ss
\item{[30]} S. Hegyi,
            {\it Phys. Lett.\/} {\bf B318} (1993) 642.\ss
\item{[31]} S. White, {\it Mon. Not. R. Astron. Soc.\/}
            {\bf 186} (1979) 145.\ss
\item{[32]} J.N. Fry,
            {\it Ap. J.\/} {\bf 306} (1986) 358.\ss
\item{[33]} Yu.L. Dokshitzer and I.M. Dremin,
            {\it Nucl. Phys.\/} {\bf B402} (1993) 139.\ss
\item{[34]} R. Peschanski, {\it ``Soft Physics and Intermittency:
            \hfill\break Open Question(s) in Krakow''\/},
            Saclay Report, Saclay-T93/118.
\item{[35]} E.A. De Wolf, private communication.

\bye